\documentclass[
    ,final            
  ]
  {aipproc}

\layoutstyle{6x9}

\begin{document}

\title{Do  Bose metals exist in Nature? }


\author{Sandro Sorella}{
  address={SISSA, INFM-Democritos,  Via Beirut n.2,  34014 Trieste,  Italy }
}

\begin{abstract}
We revisit the concept of superfluidity  
in bosonic lattice models in low dimensions.
Then,  by using  
 numerical and analytical results obtained previously 
 for equivalent spinless fermion models,  
we show that 
the  gapless phase of   1D interacting  bosons  may be 
either superfluid or -remarkably- metallic and not superfluid. The latter 
 phase -the Bose metal-  should be, 
according to the mentioned results,  a robust and stable phase in 1D. 
In higher dimensionalities  
 we speculate on the possibility of 
a stable Bose  metallic   phase   on the verge of  a Mott transition.  
\end{abstract}

\maketitle


\section{Introduction}
In the last decades there have been a lot of numerical and theoretical 
works on  interacting Bose gas in lattice or continuous 
models.\cite{ceperley,scalettar,dmrgbose,troyer,sandvick,fisher,cazalilla} 
The recent advance in the realization of optical lattices, where bosons 
are trapped on particular lattice sites and the interaction and the 
hopping parameters can be tuned continuously, 
 has also opened a novel  possibility to understand
fundamental questions  of many-body quantum mechanics, 
that can be experimentally 
checked with high degree of reliability and reproducibility.
An important example is the realization of a Mott insulating state in 
a system with strong on site repulsion.\cite{greiner2d,greiner3d}  

In this work  we want to focus on an even more fundamental question, 
that is related to the concept of superfluidity. This 
concept  deserves some discussion and generalization when considering
 a lattice model. The following 
 discussion is not at all academic because, at present, 
lattice models can be realized with laser optical techniques, and the 
gedanken experiment we will discuss in the 
next  section can be in principle realized experimentally.

\section{The model on a ring}
\begin{figure} \label{circ}
  \includegraphics[height=.6\textheight]{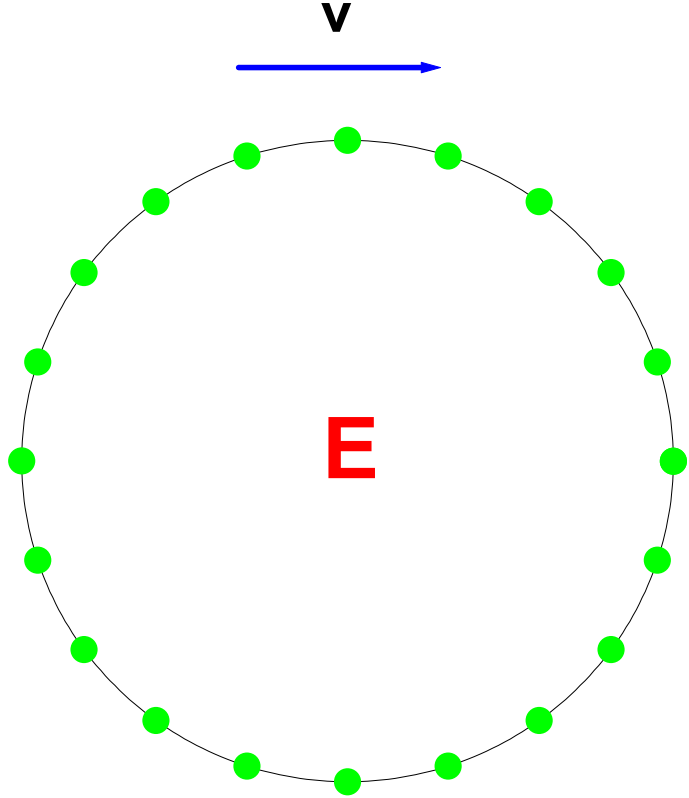}
  \caption{The model: the site positions are rotating with respect to 
the environment with given velocity $v$. The rotating 
sites are decoupled from the environment $E$, whereas the bosons can interact 
weakly with it, but are allowed to occupy only the site positions of the 
rotating ring.}
\end{figure}

We consider a one dimensional Bose-Hubbard model 
 in the  ring shown in 
Fig.(\ref{circ}). The lattice ring is rotating with given angular 
velocity $\omega_0$ with respect to the environment $E$  which is considered 
here at rest for simplicity.
Indeed in   actual experiments the environment is usually 
rotating,  but this does not change 
the forthcoming analysis, because our choice is just related to  the 
reference frame.  
 
The Hamiltonian can be generally written as:
\begin{equation}
H_v= \sum_k  \epsilon_{k} a^{\dag}_{k+v} a_{k+v}  + \hat V
\end{equation}
where $a^{\dag}_{k+v}= \sum_R e^{-i (k+v) R } a^{\dag}_R$ 
creates a boson in the ring with momentum ($\hbar =1$)
$k+v$,  $\epsilon_k$ is 
the dispersion of bosons  in the lattice (e.g. $\epsilon_k= -2t \cos k$ 
for nearest neighbor hopping), periodic  $\epsilon_{k+2 \pi}=\epsilon_k$ 
 and even $\epsilon_{k}=\epsilon_{-k}$,  
 $\hat V$ any two-body interaction term depending  only on the 
relative distance between bosons [e.g. $\hat V= U/2 \sum_R n_R (n_R-1)$ 
for the Boson-Hubbard model, where $n_R= a^{\dag}_R a_R$], thus is unaffected by the velocity $v=\omega_0 L/\pi $ of the rotating frame.  
The total momentum in  the reference system where the ring  is at rest is 
given (modulo $2 \pi$) by:
$P= \sum_k k a^{\dag}_{k+v} a_{k+v}$ and the 
momenta $k$ are obviously quantized according to the known relation 
 $k L= 2 \pi n$.
Strictly speaking in a lattice 
only the operator $e^{ i P }$ is defined, but this does not change the 
forthcoming analysis. 
In the forthcoming sections $H_0$ will be indicated by $H$ 
for simplicity.

The experimental issue to detect superfluidity is related to the following 
experiment. 
 After an  experimentally  accessible  time   (the ring rotating 
and the environment at rest)  
will  all the bosons be at rest  relative to the environment
(or  equivalently will they 
move with an appropriate velocity  with respect to the 
ring lattice positions)?
If this is not the  case we can speak about supefluidity, a fraction $\rho_s$ 
of all bosons decouple from the rest and remains  
uncoupled from the environment.

It is clear from the previous definition that superfluidity is related to the 
coupling to the environment (otherwise any finite momentum 
 will be conserved for ever in the ring). 
Nevertheless  it is possible to obtain a 
result that is independent of  the interaction between the environment and 
the ring if the following three conditions are satisfied: 
\begin{description}
\item{ i)}  the thermodynamic limit $L\to \infty$ is considered,
\item{ ii)}    a  finite temperature is given  and the low temperature 
limit is considered {\em after} that  the thermodynamic limit is employed,  
\item{iii)} the model Hamiltonian provides  a stable phase 
in the low energy spectrum, 
namely stable for small perturbation of the Hamiltonian itself.
\end{description}

The first two conditions are easily understood:
only within the finite temperature 
 canonical distribution $Z= Tr e^ { - \beta H } $ the momentum $e^{ i P }$
can equilibrate even without considering the coupling with the environment, 
and the probability of each eigenstate of the isolated ring $H$ is given 
correctly by $e^{ -\beta E_i}$,  for  a macroscopic system  ($L\to 
\infty$), just when  the coupling environment-ring is negligible with respect
 to the bulk $L$. The coupling environment-ring is used only to equilibrate 
the system and obtain a property-superfluidity- that characterizes the system 
itself and not its coupling with the environment (otherwise we could talk 
about ''superfluidity of capillary tubes'' and not superfluidity of 
e.g. $He_4$). In order to achieve this  consistent 
definition the Hamiltonian 
itself describing the system without environment 
has to define  a stable phase of matter, namely  a phase  stable 
for small  physical perturbations of the Hamiltonian, otherwise, clearly, the 
realization of a particular phase can obviously depend on the coupling 
environment-system.

In cold atoms experiments  $L$, the number of sites, 
 can be as large as $10^5$  and the thermodynamic 
limit at fixed temperature represents a realistic limit.
It is important to emphasize that  the physical 
zero temperature limit is highly non trivial in this respect.
If we take first 
$ \beta  \to \infty $  and then $ L \to \infty$ 
superfluidity {\em cannot be tested} 
because the lowest  eigenstates  of the 
Hamiltonian with non zero current  have also a non trivial complex  
momentum  $e^{i P}$  that  is obviously conserved and
 no relaxation process can occur to the real ground state in a finite size 
system.
If we explicitly consider a coupling system-environment as in\cite{landau}
to induce current relaxation, it is clear 
 that this process 
should be  essentially equivalent 
to work in the thermodynamic limit with an arbitrary small temperature. 

We conclude therefore that  the correct limit for detecting zero temperature 
superfluidity  is to take first 
 $ L\to \infty$ and then $\beta \to \infty$. This order of the  limits  
leads indeed to the definition of  superfluidity that is   independent of  
the coupling system-environment, whenever this is possible, namely  when 
(iii) is satisfied.
 
\section{Free energy and thermodynamic equilibrium}
In this section we revisit  some basic notions in thermodynamics, 
by introducing the basic quantities that define the superfluid density.
The following considerations are completely general and hold in any 
dimensionality D with minor changes, that we omit in the following.

In the thermodynamic equilibrium it is easy to show that the free energy:

\begin{equation}
F_v= -1/ \beta  \log (  Tr~e^{- \beta H_v } ) 
\end{equation}
does not depend on $v$ for discrete values of $v=v_n=2 \pi n_L /L$,  
with $n_L$ being  an  arbitrary integer.
For large size $L$ this limitation is very weak because  we can reach 
any finite  velocity $v$ for $L \to \infty$ as ,  in this limit,  the 
discrete velocity values merge in a continuum. 
Indeed for each $L$ a simple unitary transformation,  commuting with  
the two-body interaction $\hat V$:   
\begin{equation}
U= e^{ -i \sum_R  v_n  R ~ n_R } 
\end{equation}
removes  the velocity $v_n$ from $H$. 
This follows immediately after  
simple application of canonical commutation rules, implying that 
$\left[ n_{R^\prime}, a_R \right] =-\delta_{R,R^\prime} a_R$,
 so that $U^{\dag} a_{k+v} U = a_k$, and finally  
$U^{\dag} H_v U=H$ is easily obtained ($U^{\dag} a^{\dag}_{k+v} a_{k+v} U = 
(U^{\dag} a^{\dag}_{k+v} U) ( U^{\dag} a_{k+v}  U)= a^{\dag}_k a_k $ in the 
kinetic energy). 

Using the above relation,  it follows that 
the free energy:
\begin{equation}
F_v = -1/\beta \log ( Tr e^{- \beta H_v }) = -1/ \beta  
\log (Tr U^{\dag} e^{- \beta H_v } U) =-1/ \beta 
\log (  Tr e^{- \beta U^{\dag} H_v U }) =F_0
\end{equation}
does not depend on $v$ whenever $v=v_n$, namely when $U$ is defined. 
Notice that  in the first step we have used the invariance 
of the trace under cyclic permutation. 

 From this relation we can expand the partition function in  
 powers of $v$ because the mapping $k\to k+v_n$ is just a shift of the 
finite size momenta and the kinetic energy  of $H_v$ 
can be recasted in the following form: 
 $H=\sum_k \epsilon_{k-v_n} a^{\dag}_k a_{k} $ 
We thus obtain upon performing simple differentiations:
 \begin{eqnarray}
{  d F \over d v}&=&-<J>_v  + <K> v +O(v^2)=0   \label{rel} \\
J&=&   \sum_k {  d  e(k) \over d  k } a^{\dag}_k a_k \label{defj}  \\ 
K &=&   \sum_k  { d^2 e(k) \over d k^2 } a^{\dag}_k a_k  \label{defk} 
\end{eqnarray}
where the brackets  $<O>_v$ ($<O>$) 
  denote the  finite temperature averages $$ <O>_v = 
{ Tr O e^{ - \beta H_v} \over Tr e^{ - \beta H_v } }$$
 on the Hamiltonian 
of the rotating ring (non rotating ring, i.e. with  $v=0$). 
Strictly speaking the previous differentiation 
in the free energy is not allowed  because the possible velocities are 
quantized $ v_n = 2 \pi n /L$ otherwise the unitary transformation $U$ 
is not  properly  defined. 
In a more rigorous way one can indeed consider that:
\begin{equation}
 < J >_{v_n} 
 = < U^{\dag} J U > = < \sum_k { d \epsilon_{k} \over d k }|_{k+v_n}  a^{\dag}_k a_k >
\end{equation} 
In the latter equation we can 
 expand $ { d \epsilon_{k} \over d k }|_{k+v_n}= 
{  d  e(k) \over d  k } + v_n { d^2 e(k) \over d k^2 } $ and use that 
$<J>=0$ in the canonical ensemble of the rotating ring because the 
current is odd  under reflection and the  Hamiltonian $H$  is even 
for $v=0$. 
 This immediately implies the linear relation between the current
 flowing in the 
frame rotating with the ring and the corresponding 
velocity at thermal equilibrium:
\begin{equation} \label{relation}
<J >_{v_n}  = <K> v_n + O(v_n^2)
\end{equation}
within ''weak'' assumptions on the average boson occupation  in momentum space 
$n_k = a^{\dag}_k a_k $ (e.g. $1/L \sum_k { d^3 e(k) \over d k^3 } <n_k> $ 
is finite) that allows to  neglect the $O(v_n^2)$ term even for 
small but macroscopic velocities $v_n$.
It has to be remarked here that the  fundamental relation (\ref{relation}) 
 is valid only at thermal equilibrium and this may be obtained   
only  after an exceedingly large  time.
 This  is indeed the case 
 when, for $L\to \infty$, superfluidity occurs.
On the other hand whenever  
the  relation  (\ref{relation}) 
is fulfilled the current flowing in the ring is just 
representing the condition of thermal equilibrium: all the bosons by 
scattering with the environment eventually converge to an equilibrium state characterized by no   charge flow in the environment frame.

\begin{figure} \label{curr}
\includegraphics[height=.6\textheight]{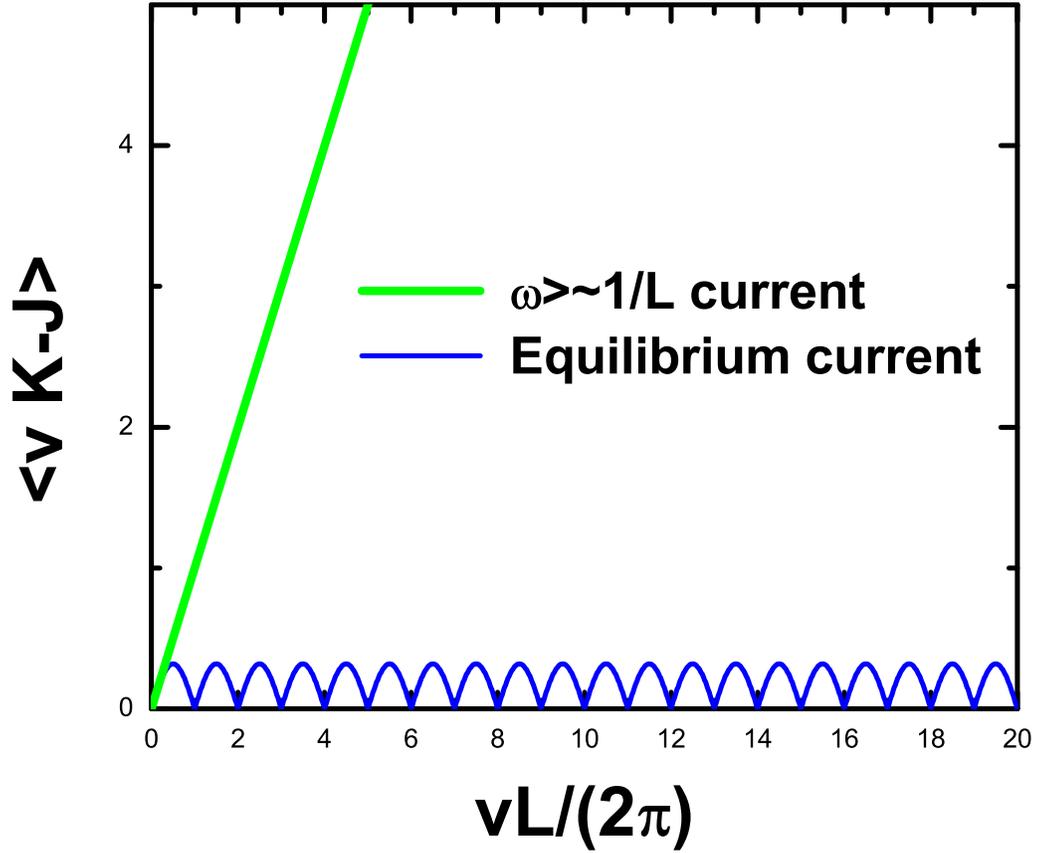}
  \caption{Equilibrium current in the environment frame as a function of 
the velocity $v$ of the rotating ring displayed in Fig.(\protect\ref{circ}). 
The  linear behavior occurs in an irrelevant 
region with exceedingly small velocity, certainly not measurable for large 
$L$, and in any case without any macroscopic current flowing in the environment 
frame.}
\end{figure}

We notice that  a linear relation between the current and the velocity 
can be obtained  within  the linear response theory.
The evaluation of $< J>_v$ for small $v$ 
is given by:
\begin{equation}
<J>_v = \left[  \int \limits_0^\beta  dt < J(t) J(0) > \right]  v 
\end{equation}
where $J(t)= e^{ t H } J e^{- t H }$. 
and  can be obtained by simple expansion of the  trace with simple and standard manipulations. 
Whenever the kernel $\alpha(0)$ relating the current response to an arbitrary 
small velocity is not equal to $<K>$ we will have a relation current velocity 
plotted in Fig(\ref{curr}). For any measurable finite velocity quantized   
as multiples of $2 \pi/L$, there is no 
net current flow  in the environment frame, implying that 
 $ < J >_v - <K> v=0(v^2)$ at equilibrium, as expected. 
However for unphysically  small values of the current the linear 
response may  have a finite  slope as shown in Fig.(\ref{curr}).   

\subsection{ Dynamical limit $\omega \to 0$}
We are arguing in the following that the 
situation displayed in Fig.(\ref{curr}) is actually the common one for 
a superfluid ($L\to \infty$ finite temperature). 
The point is that in a superfluid, in order to 
obtain the equilibrium steady state solution where no net current is 
flowing in the environment frame, an exceedingly  large time 
is necessary because an initial  current can very slowly  relax 
to the steady state.
In order to understand 
this fact it is useful to remark   that the experiments of superfluidity 
are usually done with time dependent velocity v (e.g. a torsional 
pendulum\cite{bishop}). If the frequency $\omega$ is much larger than the 
inverse relaxation time of the current 
we can safely assume that linear response theory  
can be applied  and we should obtain in this case a perfectly 
linear relation between the frequency dependent current and the 
frequency dependent velocity:
\begin{equation}
J(\omega) = \alpha(\omega) v(\omega)
\end{equation}
for either  small or  macroscopically measurable velocities as shown in the 
Fig.(\ref{curr}).
The situation is in some sense similar to the evaluation of 
the conductivity of 
a metal. The expectation value of the current in presence of a static field $E$  leads  always to zero conductivity because at 
thermal equilibrium no net current can flow. 
Indeed  we have to take the appropriate limit, with a time 
dependent field and take $\omega \to 0$ after. 
This leads to the generally accepted Kubo formula for the conductivity. 

In the superfluidity experiment, on the other hand, we have to consider the 
physical  case when the relaxation time for the current becomes macroscopically large (infinite for infinite size)   and the limit $\omega\to 0$ after the 
limit $L\to \infty$  at finite temperature $T$ leads  to:
\begin{equation}
<J>_v =\left[  \int \limits_0^\beta  dt < J(t) J(0) > \right] v= (1-\rho_s/\rho) < K > v 
\label{basic}
\end{equation}   
where $\rho_s$ is the definition of superfluid  density, being $\rho$ 
the total density of bosons. 
  Indeed 
whenever $\rho_s>0$ 
only a fraction of the particles $1-\rho_s/\rho$ 
 can be interpreted to  have  relaxed to the steady state 
  in an experimentally accessible time.
It is important to emphasize that 
 this definition of superfluidity is experimentally testable but is not 
necessarily related to broken symmetry. Indeed in two dimensions, as well known, superfluidity can be detected\cite{bishop} and on the other hand  
there is no  long range order at any  finite temperature. 
Whenever there is long range order the situation is indeed more conventional 
because the superfluid density is directly related to the helicity modulus 
of the order parameter\cite{ceperley}.  

\section{ Free bosons}
In the free boson  case,  as shown in 
\cite{ceperley},  it can be proved  that 
$\rho_s$ coincides with the condensate fraction $\rho_c$ 
of particles that occupy the $k=0$ state with a macroscopic occupation.
This calculation can be 
immediately generalized even for lattice models  
in any dimension $D$.
Since the generalization is almost immediate we describe the basic 
steps in $D=1$ for convenience of notations and write down the 
explicit expression of $\rho_s$ as a function of $D$ only in the last equation.

 For free bosons  the current $J$ commute with the Hamiltonian and the linear 
response kernel $\alpha(0)$ is given by:
\begin{equation} \label{free}
\alpha(0)= \beta < J^2 >.
\end{equation}
In fact  the Hamiltonian,   as well as $J$ (\ref{defj}),  are 
 diagonal in $k$ space $H=\sum_k \epsilon_k n_k$ $J=\sum_{k \ne 0}
  { d \epsilon_{k} \over d k } n_k$, where in the latter expression 
we have  for convenience 
removed the $k=0$ vector in the summation 
 because ${ d \epsilon_{k} \over d k }=0$ for $k=0$ (the 
derivative of an even function is odd in $k$). 
Following Ref.\cite{ceperley}, in order to evaluate Eq.(\ref{free}) 
it is enough to compute the two body density matrix in momentum space:
\begin{equation} \label{twobody}
< n_k n_{k^\prime} > = n^B_k n^B_{k^\prime} + \delta_{k,k^\prime} 
[( n^B_k)^2 +n_k^B] 
\end{equation}
where $n^B_k= 1/ [ e^( \beta (\epsilon_k -\mu) ) -1]$ is the free boson 
occupation at finite temperature, and $\mu$ is the chemical potential 
used to require a given density $\rho$ of  bosons $\rho={ 1 \over L} 
\sum_k n^B_k=\rho_c+ { 1 \over L} \sum_{k \ne 0}  n^B_k$, where $\rho_c$ 
is just the condensate density.
In this way the evaluation of $\alpha(0)$ can be readily performed and 
simplified, by using that 
i) $ \sum_{k \ne 0} n^B_k { d \epsilon_{k} \over d k }=0$ again because of the 
reflection  symmetry  ii)   as noted in Ref.\cite{ceperley} 
$[( n^B_k)^2 +n_k^B] = -1/\beta { d n^B_k \over d  \epsilon_k} $:
\begin{equation}
\alpha(0)= -\sum_{k \ne 0}  \left({ d \epsilon_k \over d k }\right) ^2
 { d n^B_k \over d \epsilon_k } 
\end{equation}
We can now take the appropriate $L \to \infty$ limit to compute $\rho_s$ 
and $\rho$ by replacing the summation $ 1/L^D \sum_{k \ne 0} 
 \to \int { dk^D  \over (2 \pi)^D} $ and 
obtain a closed form expression for $\rho_s$ (a simple integration by 
part is also left to the reader):
\begin{equation}
\rho_s/\rho =  { \rho_c \over 
\rho_c + { \int { dk^D \over (2\pi)^D } {  d^2  \epsilon_k \over dk_x^2 } 
n^B_k    \over  {  d^2  \epsilon_k \over dk_x^2 }|_{k_x=0}  } } 
\end{equation}
It is interesting that $\rho_s \ne \rho_c$ in this case, but there 
is superfluid density only when there is a  non zero condensate fraction 
and the other way around.

Thus $\rho_s=0$ for $\beta < \beta_c$ where $\beta_c$ is the inverse Bose-Einstein transition temperature that is finite in 3D but is infinite in 1D and 2D.

\subsection{Zero temperature limit}
In principle we can take first the limit $\beta \to \infty$  for the kernel 
$\alpha$ at  any fixed size $L$.
As we have emphasized before,  
this limit {\em cannot} 
 test superfluidity and indeed is related to another physical 
quantity, the zero temperature Drude weight as established by Kohn long 
time ago\cite{kohn}:
\begin{equation}
D_c =\lim_{L \to \infty} { 1\over L }  \lim_{\beta\to \infty} 
(  <K>-\alpha(0) )
\end{equation}
that distinguishes  a metal from an insulator, but not a metal from a 
superfluid.

In the following the distinction between a Bose-metal from a Bose-superfluid
is essentially analogous to the difference between a metal and  a 
superconductor valid for electronic systems\cite{scalapino}:
superconductors are obviously metal in the sense of infinite zero temperature 
conductivity but they also possess the non trivial property that the current can flow 
basically forever without dissipation at any finite temperature below $T_c$.
Clearly, within  this definition, if   $T_c=0$, we can 
speak about  a Bose-metal in the ground state because there is no measurable 
superfluid density for any $T>0$.

In order to show that the  limit  $\beta\to \infty$ 
before the thermodynamic limit 
is incorrect  for the detection 
of superfluidity, it is enough  
to realize  that, for free bosons, 
 in the limit  $\beta \to \infty$ at fixed $L$ 
the kernel $\alpha(0) \to 0$  because the current commute 
with the Hamiltonian and in the ground state $J=0$ so that 
$\alpha(0)=\beta <J^2>$ decays exponentially to zero for $\beta \to \infty$
due to the finite size gap $~1/L^2$ of the first excited state with 
non zero current.
Thus we obtain that the Drude weight for free bosons is always finite and 
equal to   $<K>$. 
Thus $D_c \ne 0$ in 1D and 2D even though at any {\em finite} temperature 
$$ <K>= \alpha(0) $$ 
for $L \to \infty$, implying $\rho_s =0$ for any $T>0$.

In our definition therefore, a free 1D or  2D 
Bose gas, is not a superfluid but a Bose 
metal. This Bose metal however is too much idealized to be considered 
realistic because interaction is always present and it is known that an 
arbitrary small interaction changes the spectrum of the excitations  from 
quadratic to linear in momentum, and condition 
iii)  for superfluidity is not satisfied. 
Thus the issue of the present paper on whether Bose metals can exist in a stable phase  is not solved by the free boson example. 
Bose metals should exist in nature only if a small physical perturbation 
of the Hamiltonian does  not change the qualitative features of the 
unperturbed phase.

Other definitions  are known in the literature for a Bose-metal\cite{doniach}, 
but  appear  much  more restrictive  definitions than the present one. 

\section{Models with interaction in 1D}
It is fortunate that the problem has been already studied  in 1D and 
we can use  convincing results obtained in other  
contexts\cite{zotos,andrei,1dbose}.
The calculation of $\rho_s$ was done with the correct order of 
limits in \cite{zotos}.  
In this work  the authors  considered 1d-spinless fermions at half filling 
with nearest neighbor hopping,  nearest $V$  and next-nearest neighbor
$W$  repulsive interactions.
This model, as well known,  is equivalent to hard-core bosons 
with the same interaction coupling 
constants, because in 1D hard core bosons with nearest neighbor hopping are 
simply related to spinless fermion models 
with the same current-current response functions. 
The spinless fermion model,  in the gapless phase,  is 
relevant for our discussion and it was clearly  found that 
indeed $\rho_s>0$ for $T>0$ as long as $W=0$ and $V>0$. 
However the authors claim that, 
an infinitesimal small coupling $W$ provides a vanishing $\rho_s$ at 
{\em any} finite temperature because-they argue- 
 the model is no longer integrable by Bethe ansatz.  
In this case the zero temperature limit of $\rho_s$ does not coincide  
with  the zero temperature Drude weight, 
that  is generally finite in 
any gapless 1D spinless fermion phase, because it is 
related to the low-energy zero-temperature properties of the model.

 If we agree with the conclusions of Ref.\cite{zotos},  
that are   based on calculations  on periodic rings with $\simeq 20$ sites, 
the model containing nearest and next-nearest neighbor repulsion 
is a Bose metal for any non zero $V,W>0$.

Indeed the conclusion of the  work\cite{zotos}   is  more general and 
, translated in the boson language, 
implies quite  generally that 1D Bose metals do  exist 
in the gapless phase. According to   the authors conjecture,   
that is still under debate, 
$\rho_s=0$ for $T>0$ in all models 
that are non integrable with Bethe ansatz (e.g. also the celebrated 
Bose-Hubbard model falls in this class if we extend this conjecture also 
to bosonic models). 
A more clear argument was given in Ref.\cite{andrei}, where the absence of 
 a finite Drude weight ($\rho_s$)  at finite temperature 
was predicted in all 1D models that do not have some  conserved current. 
Essentially, in lattice models, the current can decay due to Umklapp-processes
and a finite conductivity is expected at finite temperature, 
 a condition that is 
incompatible with a finite $\rho_s$ (which implies a $\delta-$ function 
$\omega=0$ response and therefore an infinite finite temperature conductivity). 
The condition 
of integrability may instead allow for  some conserved current, but it 
is also possible in principle 
that  some conserved current   can be realized  even  in 
non-integrable models.\cite{kawakami}  
Recently a more clear numerical evidence was also given that in a generic 
1D model with frustration $\rho_s$ is zero at finite temperature even in the 
gapless phase.\cite{1dbose} 

 We do not want to enter in this  subtle discussion  on what is the right 
criterion that allows a finite $\rho_s$ at finite temperature, but from 
what is known so far, it appears 
that only very particular  lattice models 
obtained with fine tuning of coupling constants  
 can represent  1D Bose superfluid and that the 
generic gapless phase is instead a Bose metal, at least for 
hard core boson models.   
Moreover, in this case,  the superfluid phase  obtained 
 at particular coupling strengths  do not certainly 
satisfy property (iii) and superfluidity may be detected  
only for suitable  and very particular environment-system coupling.

\section{Conclusion}
We have formulated  a consistent definition of superfluidity 
valid  for lattice and continuous  models in any  dimensionality
that relates $\rho_s$-the superfluid density-to the linear response 
 current-current correlation  calculated at finite temperature. 
This formulation agrees with the Pollock-Ceperley\cite{ceperley} 
 one based on  
the winding number, provided the correct order of limit is taken: 
first the thermodynamic limit and then the zero temperature limit, relevant  
for ground state  properties. 
In the opposite order of limits  we have shown that the so called  zero 
temperature Drude weight 
is obtained, 
but this   can be finite both 
for a  Bose-metal   and for a Bose-superfluid.
The discrimination between the two can be obtained at finite temperature  
within the present  formulation or 
by using the Scalapino-criterion 
that can be worked out directly at $T=0$\cite{scalapino}.
Both criteria coincide in the $T\to 0$ limit 
for model systems where the solution is known, 
but the latter one cannot be applied in 1D 
because it is not possible to  define  a transverse 
field in this case.

The main conclusion of our approach is that in 1D a  generic  gapless 
phase  may be  metallic and not superfluid,  
 namely a very peculiar and 
interesting interacting phase - the Bose metal- 
with {\em finite} 
 zero temperature Drude weight\cite{semimetal} but no superfluid density. 

In many recent papers the possibility to have this type of  Bose metal 
has not been considered yet, especially in 1D\cite{fisher,scalettar,cazalilla,dmrgbose}, where it has been  usually assumed that the gapless phase
 is superfluid.  This   attribute  was originally 
used to characterize the classical 2D phase corresponding 
to the 1D zero temperature quantum model. 
This  was  certainly correct but may be clearly misleading, 
because the superfluidity of the 2D classical model may be not related to 
the superfluidity of the corresponding quantum model at low temperature.

In this work we have shown that   1D hard-core boson  interacting-systems   
should be  Bose metals in the generic gapless phase, simply because 
for these models superfluidity cannot be detected   at any $T>0$ 
 (apart for the mentioned exceptions), even when the 
Drude weight is non zero in the ground state. 

Based on the above results, it appears possible  that this  
Bose metal phase can be extended also to some model without the hard
core constraint, because this constraint should not play an important 
role at low energy\cite{cazalilla}. 
The seminal work by Fisher and coworkers on the 
mapping of the 1D zero temperature Bose-Hubbard model to a classical 2D 
model at finite temperature is perfectly valid as far as the critical 
behavior at the transition 
is concerned.  However, 
 in  this mapping,  the 
superfluid density of the classical model (that can be finite below the 
 Kosterlitz-Thouless transition temperature) is related to the Drude-weight of the quantum zero temperature model and not-obviously- to its finite temperature superfluid density. 
This quantity  can be in principle  
different from the Drude weight, 
 even at arbitrary small temperature,  whenever the system is indeed 
metallic and not superfluid. It is also clear that  the analytical calculation 
of the ''superfluid density'' reported in Ref.(\cite{cazalilla}) for Luttinger 
liquids refers instead to the  zero temperature Drude weight which is obviously finite, but does 
not necessarily imply superfluidity.  On the other hand in the numerical 
calculation reported in Ref.\cite{scalettar}, no finite size scaling is attempted at fixed temperature. 
Based on these considerations it appears  important 
to improve further the numerical results of the 1D 
bosonic models  with soft or hard core constraint 
 by using recent more accurate and powerful 
 techniques\cite{sandvick}, that can be extended to much larger system sizes. 
This may allow to establish more accurately the nature of the 
gapless phases of 1D Bose models.  

In 2D close to a metal-insulator transition we have recently speculated\cite{capello} on the possibility to have a non Fermi liquid phase 
before the Mott-insulating 
phase.  
 In the boson language this possibility can be realized whenever 
the phonon velocity $c$ in the superfluid phase goes to zero before the 
Mott transition. 
In such a case an anomalous phase with finite 
zero temperature Drude weight but no superfluid density 
should appear between the Mott insulator 
and the superfluid. In this phase it can be also shown that there is 
no condensate, 
using a known relation based on the generalized indetermination 
principle.\cite{stringari} 
In the language of spin liquids the Bose-metal is just a gapless spin-liquid 
of the type stabilized in the frustrated $J_1-J_2$ model\cite{rainbow}.
Although in dimension higher than one  all these examples are clearly 
not well established because they are based on the variational approximation, 
we believe that, since in 1D the Bose (spin) liquid  is 
stable at least in hard-core boson models, it is 
worth to consider  this  phase  as a  possible phase of matter even 
in higher dimensionality and especially in 2D.  

\begin{theacknowledgments}
I acknowledge very useful discussions  with  A. Sandvik, 
R. Hlubina and A. Parola. 
 I am also especially grateful with    D. Poilblanc and 
G. Batrouni,  the organizers of the  very exciting conference in Peyres.
\end{theacknowledgments}





\begin{thebibliography}{9}
\bibitem{semimetal} Indeed the Drude weight can be also zero 
, though  with no resistivity at zero temperature as in semimetals. 
In this case it is more appropriate to 
name this phase ''Bose semimetallic phase''.
\end{thebibliography}
\end{document}